\documentclass[aps,prl,twocolumn,superscriptaddress,showpacs]{revtex4-1}

\usepackage{graphicx}
\usepackage{dcolumn}
\usepackage{bm}

\begin{document}

\title{Single-molecule spectromicroscopy: the door into
sub-diffraction refractometry}

\author{T.~A.~Anikushina}
\affiliation{Institute for Spectroscopy of Russian Academy of Sciences, Troitsk, Moscow, 142190, Russia}
\affiliation{Moscow State Pedagogical University, Moscow, 119991, Russia}
\author{M.~G.~Gladush}
\affiliation{Institute for Spectroscopy of Russian Academy of Sciences, Troitsk, Moscow, 142190, Russia}
\author{A.~A.~Gorshelev}
\affiliation{Institute for Spectroscopy of Russian Academy of Sciences, Troitsk, Moscow, 142190, Russia}
\author{A.~V.~Naumov}
\email[]{naumov@isan.troitsk.ru}
\homepage[]{www.single-molecule.ru}
\affiliation{Institute for Spectroscopy of Russian Academy of Sciences, Troitsk, Moscow, 142190, Russia}
\affiliation{Moscow State Pedagogical University, Moscow, 119991, Russia}


\begin{abstract}
 We suggest a novel approach for probing of local fluctuations of the refractive index $n$ in solids by means of single-molecule (SM) spectroscopy. It is based on the dependence $T_1(n)$ of the effective radiative lifetime $T_1$ of dye centres in solids on $n$ due to the local-field effects. Detection of SM zero-phonon lines at ultra-low temperatures gives the values of SM natural spectral linewidth (which is inverse proportional to $T_1$) and makes it possible to reveal the distribution of the local $n$ values in solids. Here we demonstrate this possibility on the example of amorphous polyethylene and polycrystalline naphthalene doped with terrylene.
\end{abstract}


\maketitle

One of the most important characteristics of a material that determine many of its macroscopic properties is the refractive index $n$. It is included as a parameter in classical equations (e.g., Fresnel's and Maxwell's), linked to dielectric permitivity and magnetic permeability, and known to depend on temperature, pressure and wavelength \cite{Ioffe:1960}. In recent years a special attention has been attracted to the refractive index in relation to metamaterials, which are assumed to have the negative $n$ \cite{PRL.90.107402,PhysRevA.76.062509}.

Since Ernst Abbe invented the refractometer in 1874 \cite{Abbe:1874} the most precise techniques for measuring $n$ have been based on the principle of the total internal reflection (TIR) on the interface of two media. There are also some other practical techniques for direct and indirect measurements of $n$ (goniometry, ellipsometry, interferometry, frustrated TIR, see, e.g. \cite{Ioffe:1960,Chamberlain:63,Muller:96,Nassif:97,Zvyagin:03}), however, most of them require averaging over a large macroscopic volume of the sample. As a result the question of the local fluctuations of $n$ in real solids remains unanswered. Moreover, in view of the growing interest to nanotechnology and nanophotonics, it is important to know whether there is a relation between $n$ and micro- and nanoscopic structure of solids. In order to give the answer some special instruments for probing the local $n$ values on the sub-diffraction (nanometer) level must be developed.
\begin{figure}[b]
\includegraphics[scale=0.40]{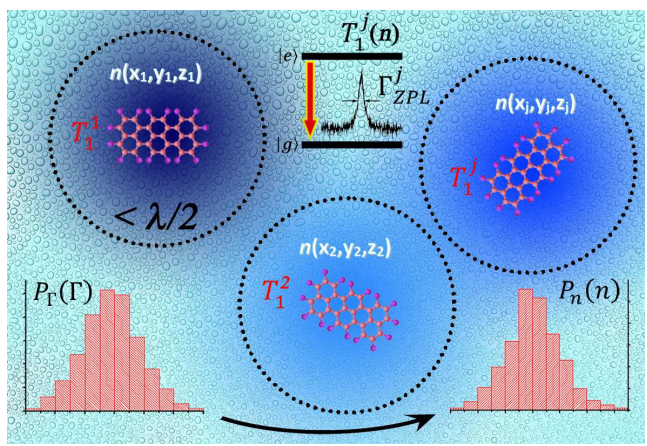}
\caption{The illustrative sketch of the developed approach for the probing of local values of the refractive index $n$ in solids. \label{fig_1}}
\end{figure}

The topical papers describe a great number of very promising ``nanoinstruments'' for probing of various local parameters of a sample based on single quantum light emitters embedded into the sample's material. For example, single-molecule (SM), single quantum dot (QD) and single nanocrystals spectroscopy and imaging provides an outstanding basis for various kind of nanoinstrumentation, like nanoscale thermometry \cite{Kucsko:2013}, nanodetectors of acoustic strain \cite{PhysRevLett.113.135505}, single charges detectors \cite{PhysRevB.90.205405}, instruments for probing of low-energy excitations in disordered solids \cite{Moerner:1997,Naumov:UFN,Osadko:book,Geva:1997,Naumov:PRLboson, Tamarat:2000}, etc. Besides, we see a possibility to use SM spectroscopy for probing the local-field effects in dye-doped solids. It is well know that the local-field effects cause the dependence of the exited state lifetime of a light emitter on the refractive index of the hosting medium $T_1(n)$ \cite{PhysRevB.60.R14012,Dolgaleva:12,Gladush:jetp2011,BarnettPRL:1992,Fleischhauer:1999,GlauberPhysRevA.43.467,PhysRevA.78.053827,PhysRevLett.81.1381}. Thus, the study of photo-physical properties of a single dye center is the way to probe the local $n$ values. For example, in \cite{Pillonnet:12} such an idea was realised by analysing the fluorescence decay of QDs.

In this work we propose a method for the probing of the local (nm-scale) fluctuations of $n$ in solids by the analysis of zero-phonon spectral lines (ZPL) of single impurity dye-molecules at ultra-low temperatures (Fig.~\ref{fig_1}).

ZPL, which corresponds to a purely electronic transition in an impurity molecule \cite{Rebane2002219}, is a unique source of information about dye-matrix interactions \cite{Moerner:1997,Naumov:UFN}. ZPL parameters (frequency, intensity, width, etc.) are very sensitive to the local environment of the corresponding chromophore SM. This fact makes SMs good candidates for spectral probes to obtain data on the structure and the internal dynamics of solids.

As we know from numerous studies, the temperature $T$ dependence of the homogeneous spectral width of ZPL is determined by three main contributions \cite{Moerner:1997,Naumov:UFN,Osadko:book,Geva:1997,Naumov:PRLboson}, i.e.:
\begin{equation}
\label{ZPL}
\Gamma_{ZPL}(T)=\Gamma_0+\Delta\Gamma_{e-tunn}(T,t_m)+\Delta\Gamma_{e-phon}(T),
\end{equation}
where the natural (lifetime limited) linewidth $\Gamma_0$ is
\begin{equation}
\label{G0}
\Gamma_0=\frac{1}{2\pi T_1};
\end{equation}
the ZPL broadening due to the interaction of electronic transitions in impurity molecules with tunneling excitations in a matrix is
\begin{equation}
\label{tunnel}
\Delta\Gamma_{e-tunn}(T,t_m)\sim T^{\alpha}\ln{(t_m)};
\end{equation}
the ZPL broadening due to the quadratic electron-phonon interaction, which in the simplest case of interaction with a single quasilocalized low-frequency vibrational mode (LFM) is expressed as
\begin{equation}
\label{phon}
\Delta\Gamma_{e-phon}(T)=w\frac{\exp(-\Delta E/kT)}{[1-\exp(-\Delta E/kT)]^2};
\end{equation}
In these equations $t_m$ is the total time of measurement, $1\leq \alpha <2$ and its exact value depends on the dye-matrix system, $w$ is the constant of quadratic electron-phonon coupling, $\Delta E$ is the energy of LFM.

ZPL width is also a function of the laser excitation intensity $P_{LAS}$:
\begin{equation}
\label{plas}
\Gamma_{ZPL}(P_{LAS})=\Gamma_{ZPL}(0)\sqrt{1+P_{LAS}/P_S},
\end{equation}
where $P_S$ is the saturation intensity and $\Gamma_{ZPL}(0)$ is the unsaturated ZPL spectral width.

From Eqs.~(\ref{ZPL})-(\ref{plas}) it follows that if $P_{LAS}\ll P_S$ and the temperature is in the range of ultra-low values, then the additional broadenings $\Delta\Gamma_{e-tunn}$ and $\Delta\Gamma_{e-phon}$ can be neglected and for each SM it is possible to measure the lifetime--limited ZPL spectral width related to $T_1$. Numerous experiments have shown, that these conditions are achieved at $P_{LAS}<0.1-10.0$ W$\cdot$cm$^{-2}$ and at $T<1-5$~$K$ (depending on the type of the matrix, see e.g. \cite{Plakhotnik199583,Navarro:14}). Thus, SMs can be used as very sensitive probes of $T_1$ fluctuations due to the local-field effects.

If we rule out the sophisticated cases and simulations of how structured environments (usually predefined by hand) may affect the lifetime of a specific light emitter (e.g., recent \cite{CPHC:CPHC200400439,PhysRevA.90.012511,Meijerink:2015}), the number of correction factors for $T_1$ used in practice can be limited to five models \cite{BarnettPRL:1992,Fleischhauer:1999,GlauberPhysRevA.43.467,PhysRevA.78.053827,Gladush:jetp2011}. They all consider the situation when a guest molecule emits light to the ``continuous medium'' of the host matrix, weakly absorbing, isotropic and characterised by $n$. Most generally, the effective $T_1$ is written as \cite{Gladush:jetp2011,Dolgaleva:12}:
\begin{equation}
\label{lifetime}
T_1(n)=\frac{\tau_0}{nf(n)},
\end{equation}
where $\tau_0$ is the ``vacuum'' value of the excited state lifetime; $n$ is the refractive index (if precisely, at the wavelength of the emitted light); function $f(n)$ reflects the local-field contributions and includes either squared or straight $n$--dependent coefficient $l(n)$ being the proportionality between the local-field acting on the emitter $E_L$ and the average Maxwellian field $E_M$, i.e., $E_L=l(n)E_M$.

To this date, the choice of $l(n)$ has been among two concepts of the local-field. One is the Lorentz local-field which is classically calculated under the assumption that the field due to polarized molecules inside a small sphere centered at the site of a light emitter may be neglected. The other is a simplified variant of Onsager's approach, \cite{Landau1984} where there is a small empty cavity around the emitter. These cases are often distinguished by the concept of interstitial and substitutional guest molecules \cite{PhysRevLett.81.1381}. The most popular models for the lifetime correction are the virtual--cavity model \cite{BarnettPRL:1992,Fleischhauer:1999} based, respectively, on the Lorentz local-field:
\begin{equation}
\label{virtual}
f(n)=\left(\frac{n^2+2}{3}\right)^2,
\end{equation}
and the real-- or empty--cavity model \cite{GlauberPhysRevA.43.467} based on the Onsager model for the local-field:
\begin{equation}
\label{real}
f(n)=\left(\frac{3n^2}{2n^2+1}\right)^2.
\end{equation}
Another approach is called the fully microscopic model \cite{PhysRevA.78.053827}, in which
\begin{equation}
\label{fullymicr}
f(n)=\frac{1}{n}\frac{n^2+2}{3}.
\end{equation}
Similar to Eq.~(\ref{virtual}) it uses the Lorentz field but here the $l(n)$ function is not squared. At the same time it misses the $n$ factor in the final form of (\ref{lifetime}). The remaining models imply $f(n)=l(n)$ \cite{Gladush:jetp2011} which brings back the missing $n$ and provides agreement with the fully microscopic model (\ref{fullymicr}) in terms that the local-field coefficients $l(n)$ are not squared. Thus, for the likely-to-be-interstitial emitters:
\begin{equation}
\label{virtual2}
f(n)=\frac{n^2+2}{3},
\end{equation}
while for the likely-to-be-substitutional emitters:
\begin{equation}
\label{real2}
f(n)=\frac{3n^2}{2n^2+1}.
\end{equation}
Speaking generally, the models in Eqs.~(\ref{virtual}) and (\ref{real}) are based on the field quantization procedures in dielectrics and other macroscopic concepts for treating the problem of spontaneous emission. The models in Eqs.~(\ref{fullymicr})-(\ref{real2}) appeared from derivation of the Maxwell-Bloch equations describing the light-matter interaction using the Heisenberg operator formalism for (\ref{fullymicr}) and Bogoliubov-Born-Green-Kirkwood-Yvon equations for reduced density matrices and correlation operators of material particles and modes of the quantized radiation field for (\ref{virtual2})-(\ref{real2}). In both approaches the field operators were initially written for material vacuum providing interactions between the guest and host particles.
\begin{figure}
\includegraphics[scale=0.31]{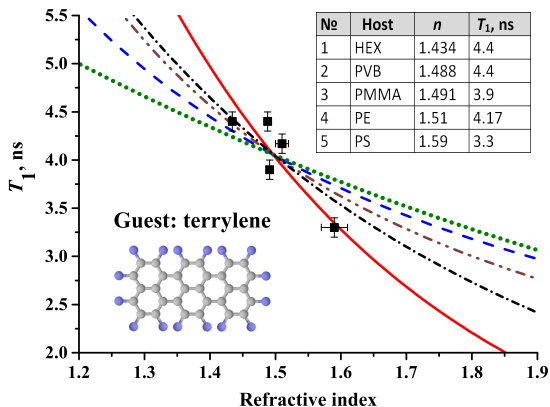}
\caption{The dependence of the excited state lifetime $T_1(n)$ of terrylene on the refractive index of its host matrix (squares). The curves show the best fits of the $T_1(n)$ dependence using different models: the virtual--cavity model (red solid); the empty--cavity model (brown dash-dot-dot); the fully microscopic models (green dot); microscopic model in Eq.~(\ref{virtual2}) (black dash-dot) and in Eq.~(\ref{real2}) (blue dashed).\label{fig_2}}
\end{figure}

Nevertheless, each of these models (\ref{virtual})-(\ref{real2}) has found ``verification'' during the analyses of the experimental data \cite{Dolgaleva:12,Gladush:jetp2011}. In all cases the so-called vacuum excitation lifetime $\tau_0$ was subjected to variation during the fitting procedure. Depending on the model used for data fitting, for each emitting system the preferable $T_1(n)$ dependence was obtained. It must be noted, however, that the verifications of the models were based on the data from the excited state lifetime measurements for dopants in crystals and glasses as well as quantum dots in various solutions. It has never been performed for such impurity systems as organic molecules in solid matrices with different degree of disorder and, specifically, has never been tried on SMs. Here we make the first attempt to test this approach using several sets of unique data obtained by different research groups.
\begin{figure}[t]
\includegraphics[scale=0.30]{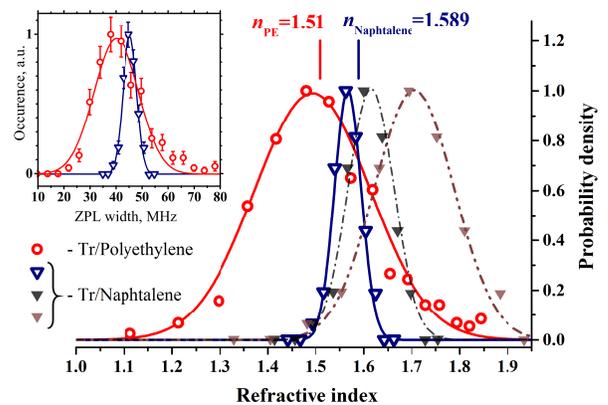}
\caption{The distribution of the local values of the refractive index in amorphous polyethylene (red circles) and polycrystalline naphthalene (blue triangles). The data are normalized to the maximum value. The sticks mark the average (for bulk media) values of the refractive index of polyethylene (red) \cite{Cooper:patent} and naphthalene (blue) \cite{CRCHandbook}. Black and brown triangles describe the distributions in naphthalene using the satisfactory fit (Eq.~(\ref{virtual2})) and the fit with Eq.~(\ref{real}). The inset shows the spectral linewidth distribution for terrylene SMs in polyethylene at 30 mK with Gaussian fit of the data \cite{Donley1999255}. The same for Tr in naphthalene crystal \cite{Donley2000109}. \label{fig_3}}
\end{figure}

In this work we have chosen terrylene (Tr) \cite{ANIEBACK:ANIE199005251} as the object of analysis. It has been widely studied in different matrixes by the laser selective spectroscopy methods, and is one of the most used fluorophores in SMS. For this compound the $T_1$ values were measured for a set of solid matrices \cite{Harms1999533}: \textit{polyethylene} (PE), \textit{polystyrene} (PS), \textit{polyvinyl-butyral} (PVB), \textit{polymethyl-methacrylate} (PMMA), and solid \textit{n-hexadecane} (Hex). In the context of the present work we also had to find the refractive indices for these matrices. The $n$ values for PE, PS, PVB were found in \cite{Cooper:patent}, whereas PMMA and HEX refractive indexes were measured by our team using the laboratory Abbe refractometer (URL-1, Russia) (see the table in Fig.~\ref{fig_2}). In Fig.~\ref{fig_2} we have plotted the $T_1$ values for Tr molecules in all of the above mentioned matrixes against the values of $n$ in these systems. The obtained $T_1(n)$ dependence was fitted using different models (Eqs.~(\ref{virtual})-(\ref{real2})). The best fit was proved to correspond to the virtual--cavity model Eq.~(\ref{virtual}) (red curve in Fig.~\ref{fig_2}), with the value of $\tau_0=12.1$  ns (i.e., $\Gamma_0=13.1$ MHz).

Given the value $\tau_0$ and the best dependence $T_1(n)$ for the Tr molecule, we can calculate the local value of $n$ at the position of a SM from its lifetime-limited ZPL width measured in the matrix involved (Fig.~\ref{fig_1}). To do this, we have taken the unique experimental data on SM ZPL width distributions form Refs.~\cite{Donley1999255,Donley2000109} obtained in the ETH, Zurich (inset in Fig.~\ref{fig_3}). The authors had performed unprecedented and complicated measurements at milliKelvin temperatures on the SMS setup equipped with the $^3$He/$^4$He dilution cryostat. At these conditions, the broadening contributions $\Delta\Gamma_{e-tunn}$ and $\Delta\Gamma_{e-phon}$ were negligible. The measurements were performed at the laser excitation intensities well below the saturation intensity. Thus, in accordance with Eq.~(\ref{plas}) the obtained data were the distributions of lifetime-limited SM ZPL widths $\Gamma_0$ directly related to SM effective excited state lifetimes $T_1$. These distributions $P_{\Gamma_0}(\Gamma_0)$ can be easily converted into the distributions of $P_{T_1}(T_1)$ using Eq.~(\ref{G0}) and then into the distributions of refraction indices $P_n(n)$ (Fig.~\ref{fig_3}) using the virtual--cavity model (Eq.~(\ref{virtual})) with the $\tau_0= 12.1$ ns.

Considering these distributions of $n$ obtained for the polycrystalline and the polymer media one should note the following: (a) there are significant fluctuations of the refractive index local values in real materials; (b) the value of $n$ measured in a bulk sample by classical methods corresponds to the peak of the $P_n(n)$ obtained in the sample. Fig.~\ref{fig_3} shows that this is valid for polyethylene and naphthalene. This implies the applicability of the proposed approach for the probing of the local fluctuations of the refractive index; (c) the local fluctuations of $n$ in the amorphous polymer are substantially greater than in the molecular polycrystal.

If one looks again at Fig.~\ref{fig_2}, another satisfactory fit is seen (black dash-dot curve), which corresponds to microscopic model (Eq.~(\ref{virtual2})). Note that it includes the Lorentz local-field factor as in the best fit with Eq.~(\ref{virtual}). The other three models result in much worse approximations. Yet, we have recalculated the distribution of lifetime-limited $\Gamma_0$ in naphthalene into the distributions of refraction indices $P_n(n)$ using all the rest models. Fig.~\ref{fig_3} shows that the distributions using the fits for Eqs.~(\ref{virtual2}) and (\ref{real}) are subjected to significant broadening and, what is most important, their peaks are notably shifted from the value of $n$ in bulk naphthalene. The other two $P_n(n)$ distributions not shown are even worse. We treat this as another evidence of the correct choice of the model and the validity of the developed approach. Note that naphthalene was absent among the systems (Fig.~\ref{fig_2}), in which $T_1$ was directly measured.

To conclude, in our study we proposed a unique approach for the probing of the local refractive index fluctuations in solids. The method is based on the detection of SM ZPLs at conditions allowing lifetime-limited spectral line widths. The great potential of this approach is demonstrated. Particularly, simultaneous reconstruction of SM spatial coordinates with the nanometre accuracy by super-resolution fluorescence microscopy \cite{NaumovEPJD:2014} opens the way to perform sub-diffraction refractometry. It was found that there are significant fluctuations of the local $n$ values in amorphous polymer and molecular polycrystalline media. These fluctuations are substantially greater in more disordered medium. The peak of the distribution $P_n(n)$ corresponds to the value of $n$, averaged over the bulk sample, which is usually obtained by traditional methods.

\begin{acknowledgments}
This work was supported by the Russian Foundation for Basic Researches: 13-02-01303 and
14-29-07270
.
\end{acknowledgments}

\bibliography{Naumovbib}

\end{document}